# Prediction of storm surge on evolving landscapes under climate change

Mohammad Ahmadi[1], David R Johnson[1,2]

[1] School of Industrial Engineering, Purdue University, West Lafayette, IN
[2] Department of Political Science, Purdue University, West Lafayette, IN

## Abstract

Planners who wish to manage coastal flood risk with long-lived infrastructure (e.g., levees, floodwalls) under a constrained computational budget face a tradeoff. Simulating a large number of future time periods or scenarios with different assumptions about land subsidence, sea level rise, land accretion, imposes a limit on how many storm simulations can be run in each scenario and time period. Machine learning approaches have been developed to reduce the computational burden of predicting storm surge from many tropical cyclone events, but prior efforts focus on predicting surge as a function of storm parameters on a single landscape. In this analysis, we present a deep learning model that also incorporates landscape characteristics and boundary conditions (e.g., mean sea level). The model is informed by a dataset of peak surge elevations from Advanced Circulation (ADCIRC) hydrodynamic simulations of coastal Louisiana in eleven scenarios: a 2020 baseline and decadal time slices from 2030 to 2070 under two scenarios varying land subsidence and sea level rise rates. Training on ten scenarios to make predictions on the eleventh yields a grand RMSE of 0.086 m and grand MAE of 0.050 m over 90 storms per scenario and over 94,000 geospatial locations. We also aggregated the 90 storms in each scenario to generate an annual exceedance probability distribution; a two-sided Kolmogorov-Smirnov test comparing AEP estimates from the model predictions to the original ADCIRC simulations rejected the null hypothesis that the predictions and ADCIRC AEP values were drawn from the same distribution only 1.1% of the time.

## Introduction

Storm surge, the temporary increase in sea level caused by severe storms, is one of the most destructive components of tropical cyclones (TC), motivating the development of accurate tools and models for its prediction (Jia et al., 2015). In recent decades, high-fidelity numerical models have been produced that are able to estimate storm surge generated by tropical cyclone wind fields with high accuracy. However, these hydrodynamic simulations typically are computationally intense, requiring high-performance computing resources, and still there can be considerable biases in their outcome (J. Zhang et al., 2018).

Because physically-based models (i.e., solving fluid dynamics equations) like the ADvanced CIRCulation (ADCIRC) model (Luettich et al., 1992) can be very expensive in terms of computational cost, surrogate models have been developed to predict storm surge without incurring the same high computational cost (Kyprioti, Taflanidis, Nadal-Caraballo, et al., 2021). The use of surrogate models or meta-models has increased rapidly in the field of coastal flood hazard research (Kyprioti et al., 2022). Surrogate models are useful tools in different fields because of their ability to emulate the behavior of complex models in their quest to approximate complex systems. Moreover, their computational efficiency makes them a convenient approach for tasks like optimization or modeling large ensembles of events and scenarios (Bartz-Beielstein & Zaefferer, 2017; Bekasiewicz et al., 2015).

Risk assessments, using techniques like joint probability methods to estimate a hazard curve, require simulation of a large number of synthetic storm events (i.e., thousands or tens of thousands). Computational limitations motivated advances such as the joint probability method with optimal sampling (JPM-OS) (Resio, 2007; Resio et al., 2009; Toro et al., 2010; Yang et al., 2019) and the use of heuristic algorithms to further reduce the number of simulations required for probabilistic flood risk (Fischbach et al., 2016; J. Zhang et al., 2018).

However, these approaches have limits, and consequently, planning studies still face meaningful constraints imposed by computational budgets. When evaluating the benefits of risk reduction infrastructure with long useful life spans, it is indeed important to consider risk and risk reduction over long planning horizons. Protection systems (e.g., levees, dikes, seawalls, pumping stations) must be designed to withstand and mitigate the effects of extreme events over many decades. This increases the necessity to consider uncertainties in factors that, over time, reshape the coastal landscape (e.g., land subsidence, land-use change, impacts of saltwater intrusion on vegetation) and boundary conditions (e.g., sea level rise) that determine risk to coastal communities; scenario analyses examine multiple future states of the world with different realizations of uncertain parameters. Integrated coastal management plans like Louisiana's *Comprehensive Master Plan for a Sustainable Coast* (Coastal Master Plan) evaluate the performance of a range of flood protection and coastal restoration projects implemented in different sequences, necessitating the modeling of multiple future time periods (Coastal Protection and Restoration Authority, 2023). Mokrech et al. (2011) stresses the importance of developing an integrated framework to assess long-term coastal impacts and thus make rational management decisions. Wamsley et al. (2009) investigated the storm surge and wave reduction benefits of different environmental restoration features (e.g., marsh restoration and barrier island changes), as well as the impact of future wetland degradation on local conditions, concluding that "consideration of natural features is required" to properly assess flood risk.

Studies that include multiple future time periods, states of the world, and project portfolios must evaluate risk on a large number of landscapes, and simple math dictates that under a fixed computational budget, the more landscapes planners want to model, the fewer events can be simulated per landscape. Using a lower-resolution mesh or a model simulating fewer physical processes may be an undesirable solution if it would introduce unacceptable biases and/or uncertainty in storm surge estimates or compromise the ability of the model to resolve the impact of protection features like levees.

In this paper, we introduce a surrogate model using artificial neural networks (ANN) that can be used to resolve this computational constraint. We train the model on synthetic storms simulated on multiple landscapes using the ADCIRC model, including not only the storm parameters as features, but also landscape features (e.g., topographic/bathymetric elevations, canopy) and boundary conditions (e.g., mean sea level). We evaluate the accuracy to predict peak storm surge elevations, as well as the accuracy of annual exceedance probability (AEP) distributions estimated using the predicted surge values, finding that the model is sufficiently accurate for use as a scenario generator in planning studies.

## Prior use of surrogate modeling for storm surge prediction

Previous studies have taken various approaches to applying surrogate models for prediction of storm surge and waves. Commonly, this means predicting peak storm surge elevations and peak significant wave heights at many points on a spatial grid as a function of the storm's characteristics at landfall. Studies vary in their choice of geography and TC characteristics, with the latter typically including parameters such as landfall location, angle of landfall, central pressure, forward velocity, radius of maximum windspeed, Holland-B parameter and/or tide level. Techniques for the surrogate models include kriging (Kyprioti, Taflanidis, Plumlee, et al., 2021; J. Zhang et al., 2018), kriging combined with principle component analysis (Jia et al., 2016; Jia & Taflanidis, 2013), support vector regression (Al Kajbaf & Bensi, 2020), and artificial neural networks (Chen et al., 2012). Al Kajbaf & Bensi (2020) provides a comparative assessment of the performance of these techniques.

Many other studies have focused on the use of surrogate models for forecasting future water surface elevations over the course of a storm (De Oliveira et al., 2009; Kim et al., 2019; Lee, 2006, 2009; Rajasekaran et al., 2008; Sztobryn, 2003). Recent works have incorporated sea level rise into predictions on a static landscape (Kyprioti, Taflanidis, Nadal-Caraballo, et al., 2021), but landscape morphology plays a significant role in modeling flood inundation and flood risk accurately (Bates et al., 2010). In areas exhibiting substantial land subsidence, erosion, barrier island migration, and other phenomena impacting morphology, incorporating sea level rise is necessary but insufficient for projecting future storm surges and inundation risks. Canopy and vegetation impact wind attenuation and surface friction, as shown in studies of mangrove forests and coastal wetlands on the Gulf coast of south Florida (K. Zhang et al., 2012). Mangrove forests reduced storm inundation areas and restricted surge inundation within a Category 3 hurricane zone, according to the study, finding that the width of the mangrove zone had a nonlinear effect on reducing surge amplitudes.

Although these previous studies investigated storm surge surrogate modeling from other perspectives, the impact of the combination of sea level rise (SLR), landscape and TC parameters on storm surge has not been thoroughly investigated. In this study, we aim to fill that gap by developing a surrogate model using deep neural networks for the prediction of peak storm surge elevations from synthetic TCs as a function of their characteristics at landfall in coastal Louisiana, four landscape parameters impacting storm surge, and mean sea level.

## Data

The synthetic tropical cyclones used in this study are characterized by their overall tracks and five parameters at landfall: forward velocity, radius of maximum windspeed, central pressure, landfall coordinates, and heading. The corpus of 645 synthetic storms was developed by the US Army Corps of Engineers for use in flood risk assessments based on the JPM-OS methodology (Nadal-Caraballo et al., 2020); each synthetic storm's landfall parameters serve as input data for the predictive model and are provided in Supplementary Information Table S1.

Hydrodynamic simulations from a coupled ADCIRC+SWAN model were available from Louisiana's 2023 Coastal Master Plan for all 645 synthetic storms, simulated on the plan's "Existing Conditions" landscape (i.e., 2020). A subset of 90 synthetic storms were simulated on each of 10 future landscapes representing decadal snapshots (i.e., 2030, 2040, 2050, 2060, 2070) under two different scenarios, a Lower and Higher Scenario, that vary in their assumptions about the rate of sea level rise, land subsidence, and other environmental factors (Coastal Protection and Restoration Authority, 2023; Cobell & Roberts, 2021). For each synthetic storm and landscape, peak storm surge elevations were extracted from the ADCIRC+SWAN simulations at 94,013 locations representing grid points from the Coastal Louisiana Risk Assessment model (CLARA) not located within fully-enclosed protection systems; the points form a mixed-resolution grid with a maximum 1-km spacing and higher resolution in some areas, such that every U.S. Census block contains at least one grid point (Johnson et al., 2023).

Each landscape is characterized by a digital elevation model defining the topography and bathymetry of the study region, as well as rasters defining other inputs to the ADCIRC model: the Manning's $n$ value (i.e., bottom roughness coefficient), free surface roughness $z0$, and a surface canopy coefficient that captures the reduction in wind stress on water surfaces produced by local vegetation. All landscape characteristics were represented as GeoTIFFs with values extracted at each of the 94,013 grid point locations for use in the surrogate model. Full details regarding the Integrated Compartment Model used to develop the landscape representations are found in White et al. (2019) and Reed and White (2023), and details regarding the ADCIRC+SWAN model and Louisiana mesh are found in Cobell and Roberts (2021) and Roberts and Cobell (2017).

## Methods

This study used feed-forward artificial neural network (ANN) models with multilayers and multiple outputs to predict storm surge at each location under current and future landscape conditions. A range of models varying from 128 neurons to 256 neurons was evaluated before selecting the models described here, as specifying too few neurons could impede the learning process while using too many neurons could result in overtraining/overfitting (Jammoussi & Ben Nasr, 2020). Moreover, for all hidden layers, the RELU activation function was chosen with a learning rate of 0.001, and for the last layer, a linear activation function was selected to predict surge values. Using 100 epochs to train the model, the entire process was executed on an AMD Epyc 7662 CPU at 2.0 GHz, taking less than 7 hours for training to be completed in preparation for making predictions on a new landscape. For all folds in the cross-validation process, it took 70 hours. Once trained, less than 4 minutes is needed to generate predictions for a novel landscape.

Firstly, we examined the value of including landscape parameters in a predictive model of storm surge for a single landscape only. ANN models were trained for current conditions on 645 synthetic storms: a multi-layered feed-forward architecture with four hidden layers and an output layer of 1 dimension was used for predicting peak storm surge elevations at different locations. A "storm-only model" at each location only included the synthetic storm parameters at landfall as inputs. The "full model" included the synthetic storm parameters but also all grid points' landscape parameters from the current landscape (latitudinal and longitudinal coordinates, topo/bathy NAVD88 elevation, surface canopy, $z0$, and Manning's $n$). Sea level was excluded from the full model in this test because the local mean sea level was assumed constant throughout the study region, and thus only has variation when multiple landscapes are taken in as input data.

Next, to investigate the impacts of climate change and the slowly evolving landscape, we trained the full model using the 2020 landscape condition and 645 synthetic storms, as well as the 10 future landscapes, each with the same 90 synthetic storms. Predictive accuracy of the full model was evaluated utilizing leave-one-out cross-validation (LOOCV) on the future landscapes; in other words, for each fold of the CV procedure, the model was trained on the 2020 landscape and 9 of the 10 future landscapes, with predictions made on the tenth future landscape. We did this to reflect a real-world use case in which the full model could serve as a scenario generator, training on a set of landscapes run through ADCIRC and then predicting outcomes in novel landscapes. The current conditions landscape's 645 storms were included to represent a realistic case in which a larger suite of synthetic TCs could be run on a single landscape as an input to a storm selection process that would identify the subset of 90 storms to run on other landscapes.

Finally, we also wanted to know how errors in the predicted peak storm surge from each synthetic storm propagate to differences in the estimated annual probability distribution of experiencing storm surge of varying elevations. This is ultimately what planners may care about when making decisions about flood protection projects. For this task, we employed the Coastal Louisiana Risk Assessment (CLARA) model, an implementation of JPM-OS which is the model used to estimate flood hazard for Louisiana's Coastal Master Plan (Johnson et al., 2013, 2023). Full details on the CLARA model's methodology are in Johnson et al. (2023); in this analysis, we compared peak surge elevation exceedance curves (i.e., surge elevations as a function of annual exceedance probability) generated from the simulated surge elevations from ADCIRC to exceedance curves generated from the predicted surge elevations from the LOOCV procedure. The resulting empirical distributions were compared using a two-sample Kolmogorov-Smirnov (KS) test, which calculates the maximum difference between two empirical samples' cumulative distribution functions to test a null hypothesis that they have been drawn from the same underlying probability distribution function (Smirnov, 1939). CLARA produces estimates of surge exceedances at 23 return periods ranging from a 50% annual exceedance probability (AEP) to 0.005% AEP (i.e., the 2-year event to the 2,000-year event), so the two-sample KS test dictates that the null hypothesis be rejected at significance level $\alpha$ if

$$\sup_x |F_{ADCIRC}(x) - F_{ANN}(x)| > \sqrt{-\ln\left(\frac{\alpha}{2}\right) \cdot \frac{1}{23}}$$

where $F_{ADCIRC}(x)$ and $F_{ANN}(x)$ are the sample CDFs associated with the ADCIRC simulations and ANN predictions, respectively.

## Results

The ANN model that includes landscape parameters performs markedly better than the model with only storm parameters when predicting surge from relatively intense storms, as shown for two illustrative storms in Figure 1. Each pane plots ADCIRC-simulated values against the ANN-predicted values at approximately 3,500 points on a west-to-east transect at 29.8° N, a latitude selected for its nearly continuous series of grid points uninterrupted by major water bodies or enclosed polders[1]. Blue points represent the Full Model which includes landscape parameters, and red points represent the Storm-Only Model which excludes them. The left-hand pane shows Storm 495, a weaker TC with central pressure of 975 mb at landfall, while the right-hand pane shows the much stronger storm 11 with a landfalling central pressure of 905 mb.

Across all 645 synthetic storms and grid points in the current conditions landscape, the Storm-only Model reached an overall RMSE of 0.31 m, while the Full Model achieved an RMSE of 0.28 m (Table 1). While this does represent an improvement of over 10 percent, primarily the result of greater accuracy for larger surge elevations, we expected the difference between these models to be minimal when trained only on the current conditions landscape. This is because of the lack of variation in landscape parameters over the synthetic storms at each point and contrastingly greater variation in TC parameters.

---

[1] The sudden decrease in points with simulated surge below 0.36 NAVD88 m is due to this being the mean sea level assumed for the current conditions landscape. Grid points over water are initialized at this level, leaving very few points along the chosen transect with lower topographic elevation, typically due to being pumped and drained.

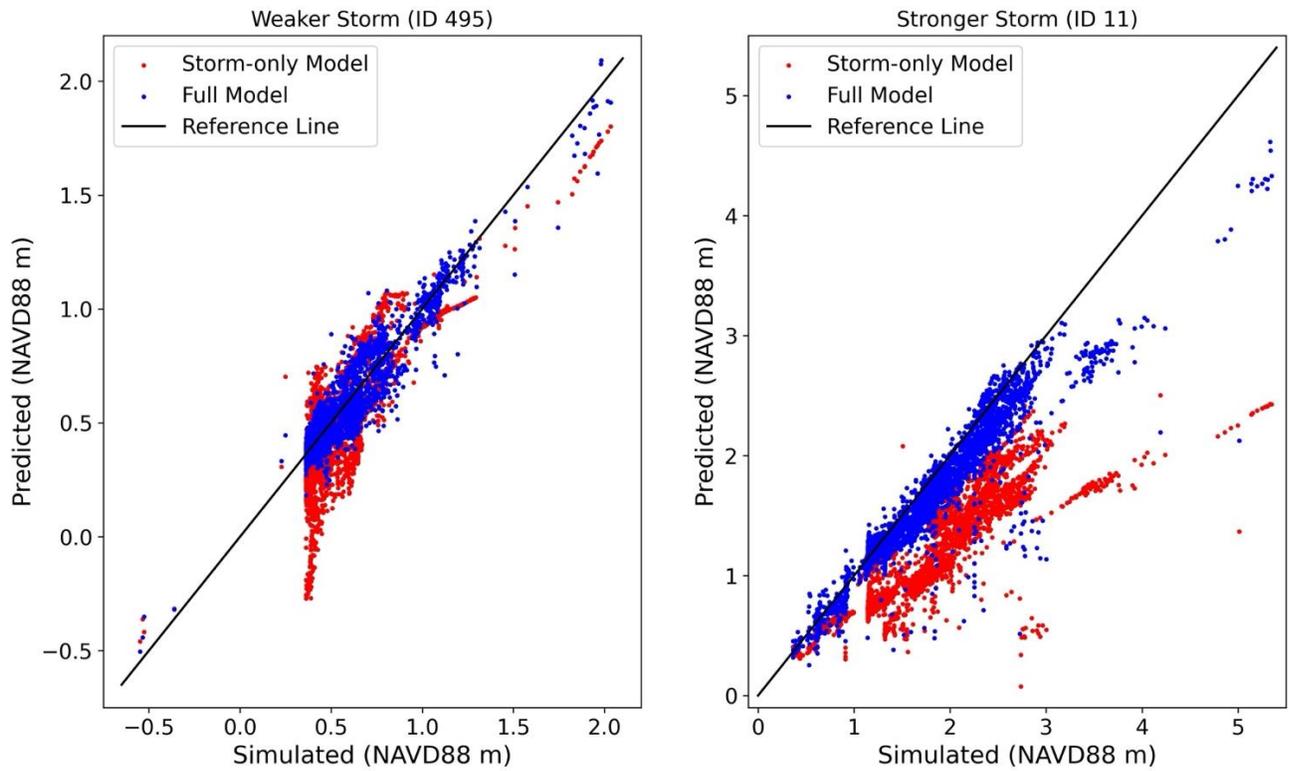

Figure 1. Simulated versus predicted storm surge for storm 495 (left pane) and storm 11 (right pane) in the cases where the ANN model input includes only storm parameters (red) and both storm and landscape parameters (blue) grid points along a transect at 29.8° N. *Note: Axis ranges vary between left and right panes.*

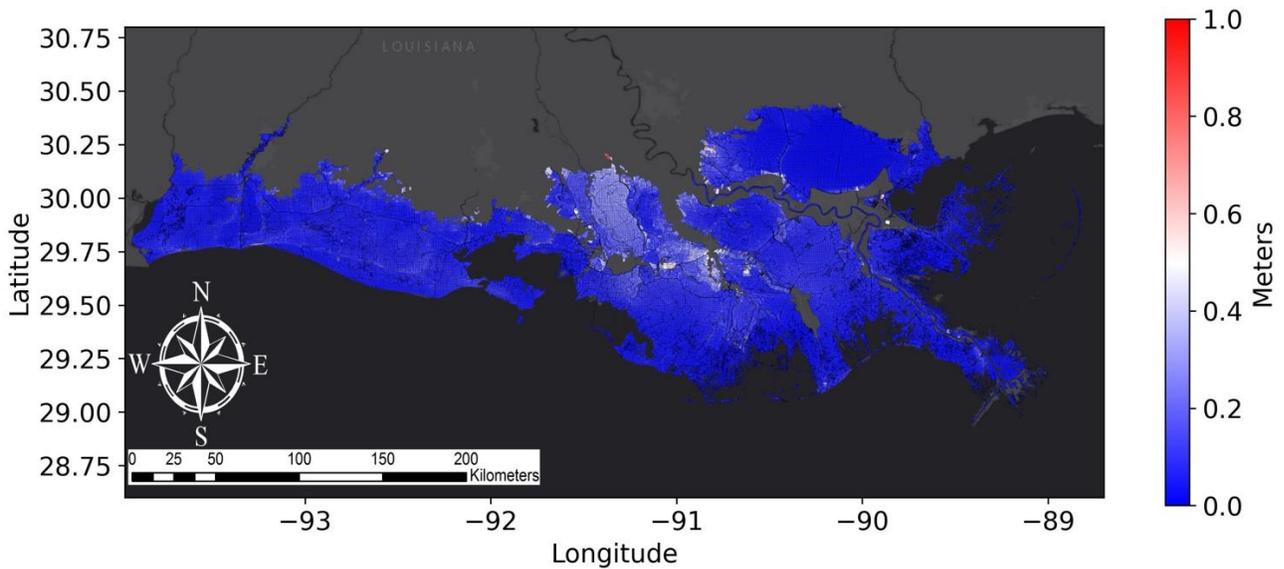

Figure 2. RMSE values across all landscapes and synthetic storms for all grid points in the study domain.

Examining the spatial pattern of the Full Model when trained on current and future scenarios, we see that points with higher RMSE over all storms and landscapes are generally further inland (Figure 2). This is expected, given that such points generally have fewer storms in the corpus that produce wetting, and we did not employ any dry-node correction techniques like those used in Shisler & Johnson (2020) or Kyprioti, et al. (2021); instead, non-wetting observations were simply removed from the training set. The model also performed less accurately in areas with more complex hydrology, such as in unpopulated wetlands in the Atchafalaya River Basin (between 91° and 92° W longitude on the northern portion of the model domain), where the ADCIRC model also has greater uncertainty and bias when validated against historic TCs (Roberts & Cobell, 2017).

Considering the RMSE of the Full Model averaged over all landscapes and synthetic storms, the RMSE at 90% of grid points is less than 0.18 m, at 99% of grid points less than 0.38 m, and at 99.9% of grid points less than 0.79 m (Figure 3). Over the ten future scenarios used in the leave-one-landscape-out cross-validation procedure, the Full Model produced a grand RMSE of 0.086 m and grand mean absolute error (MAE) of 0.050 m (Table 1). These results compare favorably to the calibration and validation results from the ADCIRC+SWAN model used to generate the hydrodynamic simulations, which reported a standard error in simulated high-water marks of approximately 0.46 m over seven historical storms (hurricanes Katrina, Rita, Gustav, Ike, Isaac, Nate, and Harvey) (Cobell & Roberts, 2021). Further analysis incorporated into the 2023 Coastal Master Plan estimated an average standard error of 0.15 m in peak surge elevations over the grid cells included in this analysis (authors' own calculations).

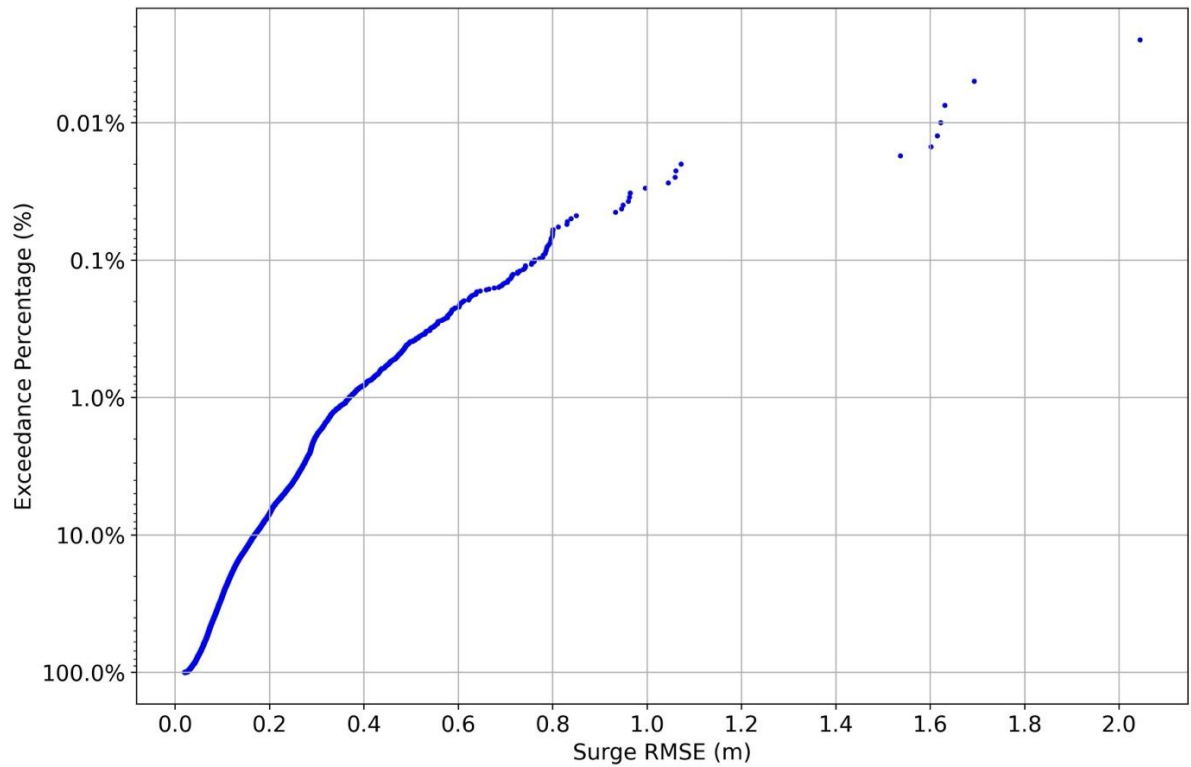

Figure 3. Exceedance percentage of RMSE values by grid point, with RMSE averaged across all landscapes and synthetic storms.

Figure 4 further disaggregates the model predictions to show the frequency distribution of predicted versus simulated surge elevations in each of the future landscapes over the synthetic storms and grid points. The overall distributions appear nearly indistinguishable except in the Higher Scenario's 2070 landscape, the most extreme scenario with respect to its assumptions about mean sea level and cumulative land subsidence. That this scenario would be an outlier compared to the others is intuitive, given its more extreme assumptions about environmental conditions; in this sense, the 2070 Higher Scenario landscape is subject to the common difficulty of extrapolating beyond training data in the leave-one-landscape-out experimental design. That said, the directionality of the difference is somewhat counterintuitive. In this scenario, the predicted storm surge is on average greater than the simulated values, though the primary non-linear difference in the scenario is an accelerating rate of sea level rise. Despite this acceleration, it appears that the Full Model overestimates the gradient in storm surge associated with changes in mean sea levels.

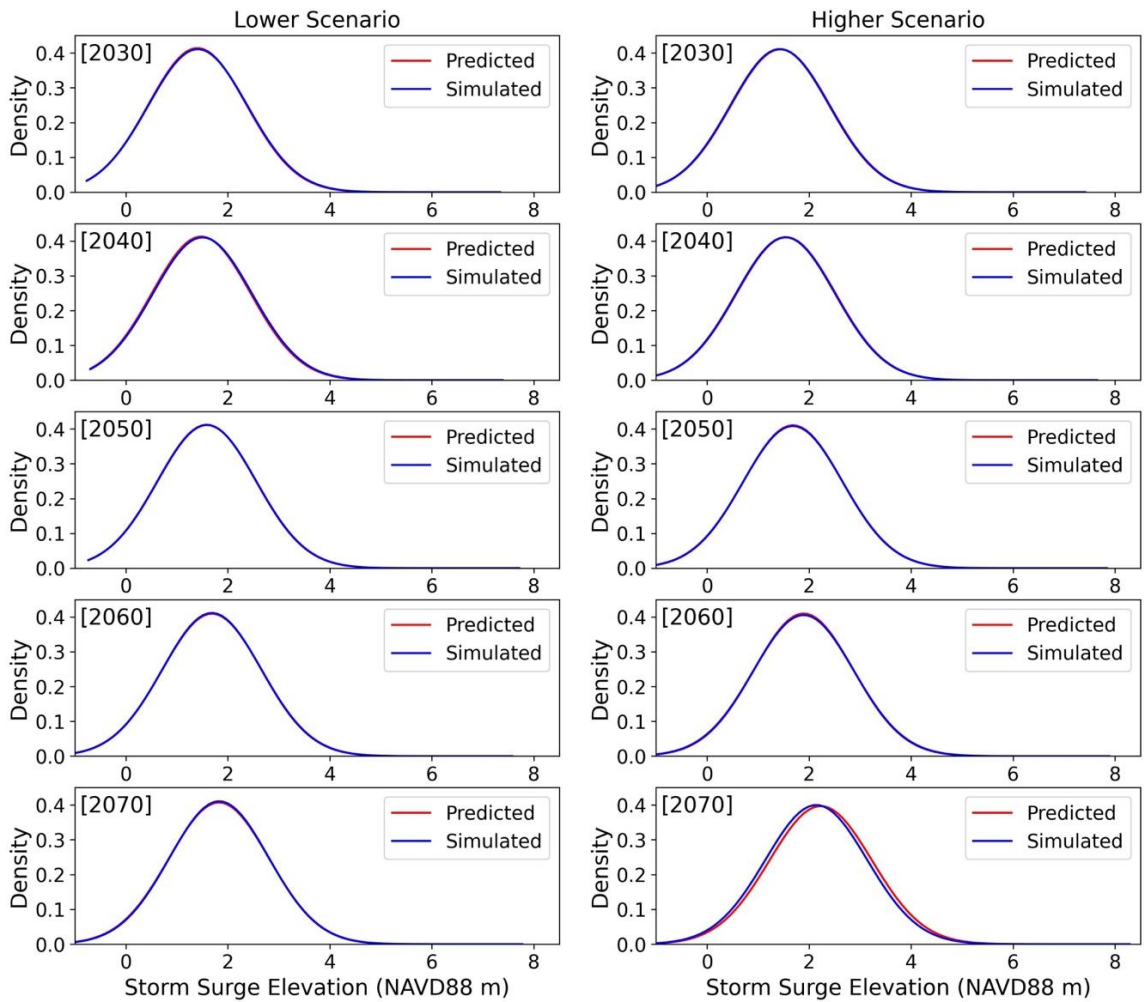

Figure 4. Frequency distribution of peak storm surge elevation (predicted versus simulated) for all future landscapes.

Examining the hazard aggregated over multiple TCs, the errors associated with surge predictions do not appear to meaningfully compound once aggregated to annual exceedance probability curves, in the sense that the RMSEs over all grid points at a range of return periods are in a similar range to the RMSEs over all grid points and synthetic storms (between 0.05 and 0.1 m for all landscapes but the most extreme, as shown in Figure 5). The RMSE generally is larger at lower AEPs, consistent with an intuition that prediction is more challenging for extreme events associated with storm surge values near the upper bounds of observations in the simulated training sets.

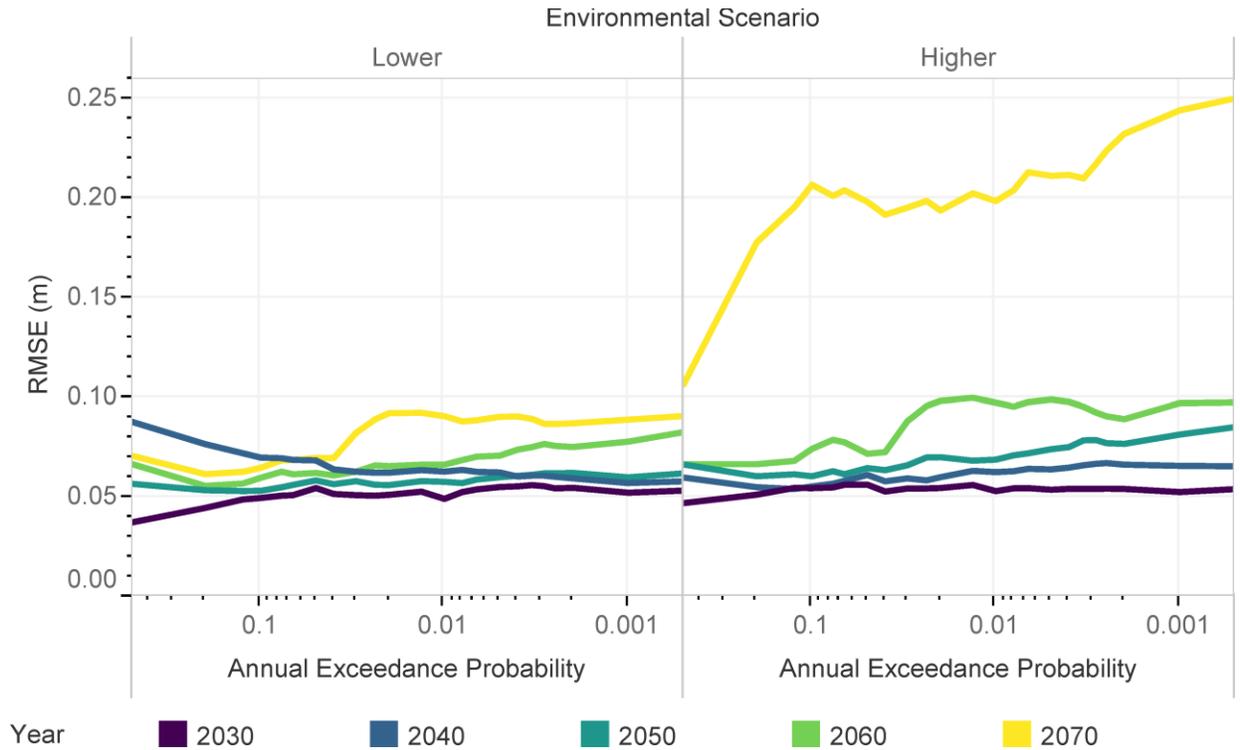

Figure 5. RMSE over all grid points, by annual exceedance probability and landscape.

Considering the full AEP distribution of storm surge at each point and in each landscape, the two-sample Kolmogorov-Smirnov tests further indicate that the surge predictions are accurate enough to usefully inform probabilistic risk studies. In eight of the ten future landscapes, the null hypothesis, that the empirical distributions generated with the ADCIRC simulations and the ANN predictions are drawn from the same underlying probability distribution, is rejected at level $\alpha = 0.05$ for less than one percent of the grid points (Table 1). This table also reports a mean absolute error (MAE) over the grid points below 0.05 m and correlation between simulated and predicted values over 0.99 for all landscapes but the 2070 Higher Scenario.

Table *1*. Summary of statistical outcomes for all cases evaluated.

| Scenarios | Years | RMSE (m) | MAE (m) | Correlation | Rejected % |
|---|---|---|---|---|---|
| Storm-only Model | 2020 | 0.314 | 0.172 | 0.912 | - |
| Full Model | 2020 | 0.277 | 0.077 | 0.965 | - |
| Higher Scenario | 2030 | 0.063 | 0.036 | 0.998 | 0.76% |
| | 2040 | 0.069 | 0.037 | 0.998 | 0.69% |
| | 2050 | 0.076 | 0.041 | 0.997 | 0.56% |
| | 2060 | 0.082 | 0.047 | 0.997 | 0.40% |
| | 2070 | 0.206 | 0.134 | 0.983 | 4.84% |
| Lower Scenario | 2030 | 0.057 | 0.035 | 0.998 | 0.74% |
| | 2040 | 0.089 | 0.044 | 0.996 | 1.43% |
| | 2050 | 0.064 | 0.037 | 0.998 | 0.46% |
| | 2060 | 0.073 | 0.040 | 0.997 | 0.53% |
| | 2070 | 0.081 | 0.045 | 0.997 | 0.81% |

From Table 1, it is evident that by incorporating landscape parameters into the ANN model, storm surge can be predicted accurately for a variety of different scenarios. Figure 6 highlights the spatial pattern of points that rejected the null hypothesis of the two-sided K-S test for an illustrative landscape, the Higher Scenario in 2060. Red points indicate the locations where the test rejects the null hypothesis at $\alpha = 0.05$, while the blue points indicate locations where the evidence fails to rule out the possibility of the hazard estimates coming from the same underlying AEP distribution.

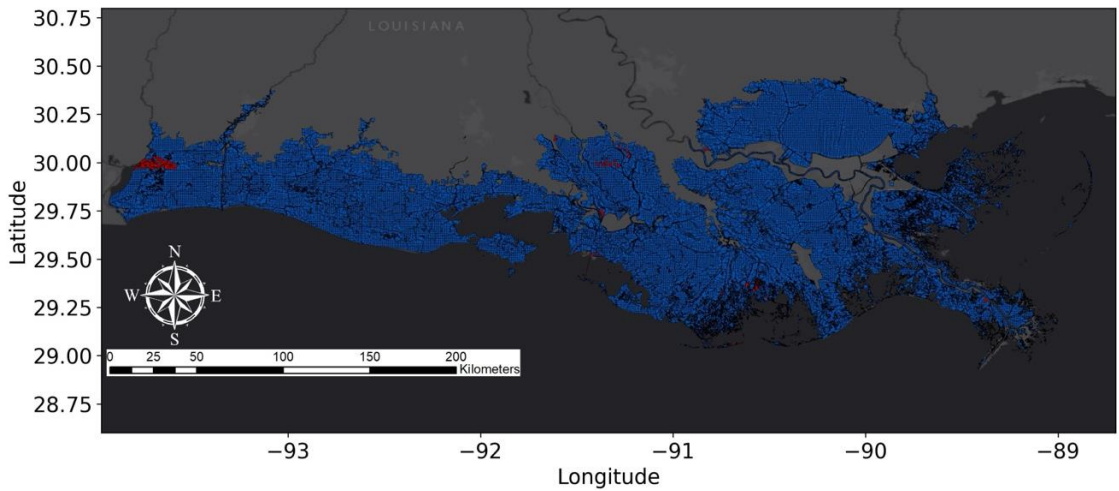

Figure 6. Results of two-sample KS test for the year 2060 of the Higher Scenario. Red indicates points where the null hypothesis is rejected.

## Discussion and Conclusions

We have presented a machine learning-based surrogate model of peak storm surge elevations that yields predictions of comparable or greater accuracy, when compared to ADCIRC simulations, than the ADCIRC model relative to historic observations used for model calibration and validation. The addition of future landscapes with variation in landscape parameters and mean sea level conditions provides more features and heterogeneity in training data to improve the model's accuracy. In a leave-one-landscape-out cross-validation exercise, the model produced a mean absolute error of approximately 0.04 m in nine out of ten of the future landscapes, with the exception being the year 2070 of the Higher Scenario, the most extreme landscape with respect to having the greatest sea level rise and land subsidence. A two-sided Kolmogorov-Smirnov test failed to reject the null hypothesis, that points on hazard curves generated by the ADCIRC simulations and ANN predictions are drawn from the same underlying distribution, at less than 1% of grid points in eight out of ten of the future landscapes.

This highlights an important caveat: this analysis utilized ADCIRC simulations that were readily available from Louisiana's 2023 Coastal Master Plan, meaning that the scenarios and time periods were not chosen with the idea of using ADCIRC to train a surrogate model already in mind. This work therefore represents a proof of concept where the ANN model produced predictions suitable for planning studies from a training set of convenience. If planners are interested in estimating risk over a 50-year planning horizon ending in 2070, it may be that accuracy could be improved for the same computational cost by replacing one of the "intermediate" landscapes with a landscape corresponding to the year 2080 instead, to mitigate the challenges of ML models in extrapolating beyond data in their training set.

This also has implications for the storm selection process, given that the 90 storms simulated in each future landscape were chosen by comparing hazard curves to the curve associated with the full 645-storm suite in the current conditions landscape (Fischbach et al., 2021). Prior research has suggested a difficulty in using ML methods to predict extreme storm surge elevations (Hashemi et al., 2016; Ramos-Valle et al., 2021). While accurate reproduction of extreme individual events is important in controlling the overall RMSE and MAE of the model, extreme storms (i.e., with lower central pressures at landfall) are relatively more rare in occurrence, thus having smaller probability masses when contributing to annual exceedance probability distributions and making smaller contributions to expected annual damage calculations.

Consequently, adoption of surrogate models as scenario generators would also benefit from a rigorous consideration of how optimal sampling techniques could be extended to include heterogeneity in landscape parameters and boundary conditions. When planning an analysis that will span a range of future states of the world, it is likely that greater computational efficiency could be achieved by sampling different synthetic storm events on each landscape, rather than simulating the same 90 storms as was done for the Coastal Master Plan.

All the future landscapes used for training our ANN came from the Coastal Master Plan's Future Without Action scenarios (i.e., no additional projects implemented on the landscape). This means we have restricted our predictions to analyzing a slowly evolving landscape without major coastal management interventions. However, the surrogate model developed in this study would also have utility in evaluating the flood risk impacts of coastal restoration projects that affect landscape morphology over time scales ranging from the immediate (e.g., beach nourishment) to decadal (e.g., river diversions).

This study enables better modeling of future climate and environmental conditions by policy makers and water resource managers. Moreover, the developed model makes it possible to evaluate risk under a greater number and range of future scenarios and time periods, opening the door to the use of computationally expensive models like ADCIRC for planning studies utilizing techniques for decision-making under deep uncertainty that require or benefit from the use of large ensembles of future states of the world (Johnson & Geldner, 2019).

## Author contributions

David Johnson designed the conceptual framework and analysis plan and led the writing of the paper. Mohammad Ahmadi developed, debugged, and refined the model, executed the analysis, and contributed to writing the paper.

## Data and code availability

Model source code, written in Python 3 and R, is available at https://github.com/mohammadahmadi1995/Storm_Surge. Data for this project are available at DesignSafe-CI, https://doi.org/10.17603/ds2-0ksb-yy40.

## Acknowledgements

This work was supported by the U.S. National Science Foundation under awards 2238060 and 2118329. We thank members of the modeling teams for Louisiana's *Comprehensive Master Plan for a Sustainable Coast* and funding from the Coastal Protection and Restoration Authority for the original simulation data used in this study. Any opinions, findings, conclusions, and recommendations expressed in this material are those of the authors and do not necessarily reflect the views of the funding entities.

# Supplementary Information

Table S1. Distribution of synthetic storm parameters at landfall.

| Storm ID | Heading | $v_f$ (knots) | $r_{max}$ (nm) | Landfall lon ($x$) | $c_p$ (mbar) |
|---|---|---|---|---|---|
| 1 | 35.8 | 9.5 | 10.9 | -102.376 | 865.25 |
| 2 | 35.8 | 15.3 | 14.7 | -102.375 | 885.25 |
| 3 | 35.8 | 11.8 | 5.9 | -102.377 | 905.25 |
| 4 | 35.8 | 9.1 | 9.2 | -102.377 | 925.25 |
| 5 | 35.8 | 16.7 | 15.5 | -102.378 | 945.25 |
| 6 | 35.8 | 8.3 | 31.3 | -102.376 | 965.25 |
| 7 | 35.8 | 10.2 | 59 | -102.377 | 985.25 |
| 8 | 35.8 | 9.9 | 23.4 | -102.377 | 1005.25 |
| 9 | 62.72727 | 20.6 | 9 | -98.8967 | 865.25 |
| 10 | 62.72727 | 7.3 | 5.1 | -98.8991 | 885.25 |
| 11 | 62.72727 | 8.6 | 27.3 | -98.8964 | 905.25 |
| 12 | 62.72727 | 10.2 | 25.3 | -98.8974 | 925.25 |
| 13 | 62.72727 | 4.8 | 27.5 | -98.8975 | 945.25 |
| 14 | 62.72727 | 9.3 | 22.4 | -98.8986 | 965.25 |
| 15 | 62.72727 | 9.7 | 11.7 | -98.899 | 985.25 |
| 16 | 62.72727 | 10.9 | 44.5 | -98.899 | 1005.25 |
| 17 | 69.86364 | 9.8 | 5 | -95.3612 | 865.25 |
| 18 | 69.86364 | 14.4 | 11.9 | -95.3631 | 885.25 |
| 19 | 69.86364 | 5.1 | 16.4 | -95.36 | 905.25 |
| 20 | 69.86364 | 17.2 | 10.2 | -95.359 | 925.25 |
| 21 | 69.86364 | 7.8 | 36.8 | -95.3605 | 945.25 |
| 22 | 69.86364 | 9.8 | 25.1 | -95.3612 | 965.25 |
| 23 | 69.86364 | 4.6 | 9 | -95.3626 | 985.25 |
| 24 | 69.86364 | 12.3 | 56.6 | -95.3634 | 1005.25 |
| 25 | 69.88333 | 10.9 | 6 | -91.8512 | 865.25 |
| 26 | 69.88333 | 5.2 | 8 | -91.85 | 885.25 |
| 27 | 69.88333 | 10.5 | 19.8 | -91.85 | 905.25 |
| 28 | 69.88333 | 6.7 | 42.3 | -91.8499 | 925.25 |
| 29 | 69.88333 | 17.5 | 26.5 | -91.8528 | 945.25 |
| 30 | 69.88333 | 8.6 | 11.4 | -91.8494 | 965.25 |
| 31 | 69.88333 | 11.8 | 51.2 | -91.8525 | 985.25 |
| 32 | 69.88333 | 5.2 | 35.9 | -91.8503 | 1005.25 |
| 33 | 77.34848 | 11.2 | 7.7 | -88.3561 | 865.25 |
| 34 | 77.34848 | 18.1 | 15.7 | -88.3545 | 885.25 |
| 35 | 77.34848 | 11.4 | 17.9 | -88.3557 | 905.25 |
| 36 | 77.34848 | 8.5 | 11.8 | -88.3535 | 925.25 |
| 37 | 77.34848 | 9.3 | 49.6 | -88.3521 | 945.25 |
| 38 | 77.34848 | 5 | 27.1 | -88.3541 | 965.25 |
| 39 | 77.34848 | 4.5 | 19 | -88.3534 | 985.25 |
| 40 | 77.34848 | 13.2 | 33.4 | -88.355 | 1005.25 |
| 41 | 88.96364 | 6.3 | 8.8 | -85.1718 | 865.25 |
| 42 | 89.01786 | 11.3 | 6.3 | -85.1751 | 885.25 |
| 43 | 88.96364 | 13.2 | 29 | -85.1714 | 905.25 |
| 44 | 88.96364 | 8 | 34.8 | -85.1727 | 925.25 |
| 45 | 88.96364 | 6.2 | 17.6 | -85.1724 | 945.25 |
| 46 | 88.96364 | 17 | 23.3 | -85.1735 | 965.25 |
| 47 | 88.96364 | 8.5 | 21.9 | -85.1722 | 985.25 |
| 48 | 88.96364 | 7.7 | 62.4 | -85.1726 | 1005.25 |
| 49 | 48.39024 | 9.6 | 17.7 | -96.13 | 875.25 |
| 50 | 48.39024 | 15.9 | 16.4 | -96.1307 | 895.25 |
| 51 | 48.39024 | 8.7 | 8.9 | -96.1314 | 915.25 |
| 52 | 48.39024 | 9.6 | 9.3 | -96.13 | 935.25 |
| 53 | 48.39024 | 16.8 | 13 | -96.1303 | 955.25 |
| 54 | 48.39024 | 11.5 | 50.9 | -96.131 | 975.25 |
| 55 | 48.39024 | 4.8 | 32.6 | -96.1309 | 995.25 |
| 56 | 49.06522 | 21.9 | 13.9 | -94.9388 | 875.25 |
| 57 | 49.06522 | 14.9 | 9.4 | -94.9391 | 895.25 |
| 58 | 49.06522 | 9.3 | 7.6 | -94.9392 | 915.25 |
| 59 | 49.06522 | 10.5 | 22.4 | -94.9383 | 935.25 |
| 60 | 49.06522 | 5.3 | 16 | -94.9385 | 955.25 |
| 61 | 49.06522 | 11.9 | 48.7 | -94.9389 | 975.25 |
| 62 | 49.06522 | 12 | 15 | -94.9388 | 995.25 |
| 63 | 51.03846 | 24.4 | 7.6 | -93.7258 | 875.25 |
| 64 | 51.03846 | 13.2 | 9.7 | -93.7262 | 895.25 |
| 65 | 51.03846 | 5.5 | 19.4 | -93.7253 | 915.25 |
| 66 | 51.03846 | 8 | 44.4 | -93.7255 | 935.25 |
| 67 | 51.03846 | 8.3 | 9.2 | -93.7247 | 955.25 |
| 68 | 51.03846 | 12.6 | 35.9 | -93.7253 | 975.25 |
| 69 | 51.03846 | 8.4 | 28 | -93.725 | 995.25 |
| 70 | 51.05172 | 5.3 | 13.4 | -92.5434 | 875.25 |
| 71 | 51.05172 | 17.5 | 11.8 | -92.5442 | 895.25 |
| 72 | 51.05172 | 9.6 | 9.7 | -92.5436 | 915.25 |
| 73 | 51.05172 | 4.9 | 15.4 | -92.5437 | 935.25 |
| 74 | 51.05172 | 5.8 | 27 | -92.5433 | 955.25 |
| 75 | 51.05172 | 9.8 | 44.8 | -92.5445 | 975.25 |
| 76 | 51.05172 | 6 | 11.2 | -92.5439 | 995.25 |
| 77 | 52.05085 | 13.6 | 11.2 | -91.324 | 875.25 |
| 78 | 52.05085 | 9.2 | 13.2 | -91.325 | 895.25 |
| 79 | 52.05085 | 14.5 | 20.7 | -91.3244 | 915.25 |
| 80 | 52.05085 | 9.3 | 48 | -91.3253 | 935.25 |
| 81 | 52.05085 | 9.6 | 12.2 | -91.3248 | 955.25 |
| 82 | 52.05085 | 6.2 | 28.8 | -91.324 | 975.25 |
| 83 | 52.05085 | 15 | 47 | -91.3246 | 995.25 |
| 84 | 54.83051 | 6.8 | 13 | -90.1105 | 875.25 |
| 85 | 54.83051 | 15.4 | 12.9 | -90.1105 | 895.25 |
| 86 | 54.83051 | 12.5 | 14.4 | -90.1102 | 915.25 |
| 87 | 54.83051 | 9.1 | 41.7 | -90.11 | 935.25 |
| 88 | 54.83051 | 13.6 | 22.5 | -90.1105 | 955.25 |
| 89 | 54.83051 | 5.6 | 20 | -90.1113 | 975.25 |
| 90 | 54.83051 | 11 | 55.5 | -90.1107 | 995.25 |
| 91 | 57.39063 | 8.3 | 7.8 | -88.9005 | 875.25 |
| 92 | 57.39063 | 9.4 | 7.2 | -88.9003 | 895.25 |
| 93 | 57.39063 | 8.2 | 27.6 | -88.9001 | 915.25 |
| 94 | 57.39063 | 13.2 | 14.8 | -88.9003 | 935.25 |
| 95 | 57.39063 | 9.9 | 51.2 | -88.9006 | 955.25 |
| 96 | 58.02985 | 10.9 | 14.6 | -88.9 | 975.25 |
| 97 | 58.02985 | 5.1 | 37.7 | -88.8991 | 995.25 |
| 98 | 60.03077 | 9.9 | 9.6 | -87.7248 | 875.25 |
| 99 | 60.03077 | 5.7 | 18.7 | -87.7257 | 895.25 |
| 100 | 60.03077 | 6.9 | 13.4 | -87.7242 | 915.25 |
| 101 | 60.03077 | 15 | 10.5 | -87.7249 | 935.25 |
| 102 | 60.03077 | 12.7 | 59.2 | -87.7245 | 955.25 |
| 103 | 60.03077 | 12.2 | 33.4 | -87.7247 | 975.25 |
| 104 | 60.03077 | 8.9 | 33.8 | -87.7245 | 995.25 |
| 105 | 66.86667 | 12.7 | 15 | -86.4704 | 875.25 |
| 106 | 66.86667 | 11.4 | 6.6 | -86.4712 | 895.25 |
| 107 | 66.86667 | 5.8 | 18.2 | -86.4713 | 915.25 |
| 108 | 66.86667 | 14.5 | 11.7 | -86.4705 | 935.25 |
| 109 | 66.86667 | 10.7 | 54.6 | -86.4709 | 955.25 |
| 110 | 66.86667 | 13.4 | 23.7 | -86.4713 | 975.25 |
| 111 | 66.86667 | 9.3 | 69.1 | -86.4709 | 995.25 |
| 112 | 69.01786 | 8.6 | 6.4 | -85.25 | 875.25 |
| 113 | 69.01786 | 10.7 | 20.1 | -85.2502 | 895.25 |
| 114 | 69.01786 | 9 | 10.6 | -85.2499 | 915.25 |
| 115 | 69.01786 | 16 | 13.6 | -85.25 | 935.25 |
| 116 | 69.01786 | 5.6 | 19.1 | -85.251 | 955.25 |
| 117 | 69.01786 | 9.2 | 68.6 | -85.2507 | 975.25 |
| 118 | 69.01786 | 7.3 | 22.8 | -85.2501 | 995.25 |
| 119 | 29.62791 | 7.9 | 9.2 | -95.5716 | 865.25 |
| 120 | 29.62791 | 21.2 | 9.8 | -95.5717 | 885.25 |
| 121 | 29.62791 | 13.7 | 8 | -95.5719 | 905.25 |
| 122 | 29.62791 | 9.9 | 33.2 | -95.5718 | 925.25 |
| 123 | 29.62791 | 6.9 | 18.4 | -95.5716 | 945.25 |
| 124 | 29.62791 | 15.6 | 9 | -95.5719 | 965.25 |
| 125 | 29.62791 | 9.2 | 36.3 | -95.5716 | 985.25 |
| 126 | 29.62791 | 4.6 | 15.4 | -95.5718 | 1005.25 |
| 127 | 29.06522 | 8.5 | 5.6 | -94.7963 | 865.25 |
| 128 | 29.06522 | 23.7 | 16.8 | -94.7964 | 885.25 |
| 129 | 29.06522 | 8.1 | 10.7 | -94.7971 | 905.25 |
| 130 | 29.06522 | 11.8 | 6.6 | -94.7967 | 925.25 |
| 131 | 29.06522 | 19.4 | 32.8 | -94.7964 | 945.25 |
| 132 | 29.06522 | 5.5 | 40.4 | -94.7961 | 965.25 |
| 133 | 29.06522 | 10.8 | 26 | -94.7964 | 985.25 |
| 134 | 29.06522 | 4.3 | 12.6 | -94.7964 | 1005.25 |
| 135 | 28.97959 | 16.2 | 6.2 | -94.0161 | 865.25 |
| 136 | 28.97959 | 16.3 | 5.6 | -94.016 | 885.25 |
| 137 | 28.97959 | 15.4 | 11.4 | -94.0157 | 905.25 |
| 138 | 28.97959 | 8.3 | 29.2 | -94.0159 | 925.25 |
| 139 | 28.97959 | 5.7 | 35.4 | -94.0148 | 945.25 |
| 140 | 28.97959 | 4.9 | 13 | -94.0158 | 965.25 |
| 141 | 28.97959 | 15.7 | 31.4 | -94.0163 | 985.25 |
| 142 | 28.97959 | 9 | 14.5 | -94.0163 | 1005.25 |
| 143 | 30.55172 | 23.7 | 4.6 | -93.2425 | 865.25 |
| 144 | 30.55172 | 18.7 | 10.1 | -93.2417 | 885.25 |
| 145 | 30.55172 | 6.7 | 9.9 | -93.2419 | 905.25 |
| 146 | 30.55172 | 14.5 | 26.2 | -93.242 | 925.25 |
| 147 | 30.55172 | 10.8 | 9.4 | -93.2426 | 945.25 |
| 148 | 30.55172 | 6.7 | 47.6 | -93.2425 | 965.25 |
| 149 | 30.55172 | 5 | 30.3 | -93.2418 | 985.25 |
| 150 | 30.55172 | 9.2 | 20.4 | -93.2424 | 1005.25 |
| 151 | 30.48276 | 27 | 12.7 | -92.4729 | 865.25 |
| 152 | 30.48276 | 12.4 | 7.5 | -92.4726 | 885.25 |
| 153 | 30.48276 | 6.2 | 19.2 | -92.4723 | 905.25 |
| 154 | 30.48276 | 8.8 | 12.3 | -92.4727 | 925.25 |
| 155 | 30.48276 | 18.4 | 8 | -92.4726 | 945.25 |
| 156 | 30.48276 | 13.1 | 38.9 | -92.4725 | 965.25 |
| 157 | 30.48276 | 11.4 | 21 | -92.4725 | 985.25 |
| 158 | 30.48276 | 7.9 | 40 | -92.4724 | 1005.25 |
| 159 | 30.15 | 19.8 | 5.9 | -91.6849 | 865.25 |
| 160 | 30.15 | 16.8 | 20.2 | -91.6852 | 885.25 |
| 161 | 30.15 | 21 | 9.2 | -91.685 | 905.25 |
| 162 | 30.15 | 7.2 | 13.9 | -91.6848 | 925.25 |
| 163 | 30.15 | 14.5 | 7.4 | -91.6843 | 945.25 |
| 164 | 30.15 | 5.9 | 37.5 | -91.6849 | 965.25 |
| 165 | 30.15 | 13.3 | 35 | -91.6848 | 985.25 |
| 166 | 30.15 | 7.3 | 10.7 | -91.6844 | 1005.25 |
| 167 | 34.58333 | 13.9 | 13.3 | -90.8992 | 865.25 |
| 168 | 34.58333 | 9.1 | 5.3 | -90.899 | 885.25 |
| 169 | 34.58333 | 5.6 | 14.4 | -90.8989 | 905.25 |
| 170 | 34.58333 | 15.4 | 23.6 | -90.8992 | 925.25 |
| 171 | 34.58333 | 6.4 | 29.5 | -90.8993 | 945.25 |
| 172 | 34.58333 | 9.1 | 32.5 | -90.899 | 965.25 |
| 173 | 34.58333 | 11.1 | 72.2 | -90.8988 | 985.25 |
| 174 | 34.58333 | 8.5 | 16.4 | -90.8984 | 1005.25 |
| 175 | 34.83051 | 9.2 | 6.8 | -90.1173 | 865.25 |
| 176 | 34.83051 | 9.4 | 13.8 | -90.1173 | 885.25 |
| 177 | 34.83051 | 14.5 | 14 | -90.117 | 905.25 |
| 178 | 34.83051 | 7.7 | 39.1 | -90.1168 | 925.25 |
| 179 | 34.83051 | 15 | 14.8 | -90.1171 | 945.25 |
| 180 | 34.83051 | 19.3 | 29.2 | -90.1171 | 965.25 |
| 181 | 34.83051 | 9.4 | 27 | -90.1167 | 985.25 |
| 182 | 34.83051 | 9.7 | 59.3 | -90.1174 | 1005.25 |
| 183 | 35.16129 | 11.9 | 9.5 | -89.3274 | 865.25 |
| 184 | 35.16129 | 13.2 | 15.1 | -89.3277 | 885.25 |
| 185 | 35.16129 | 7.5 | 7.7 | -89.327 | 905.25 |
| 186 | 35.16129 | 5.9 | 27.1 | -89.3273 | 925.25 |
| 187 | 35.16129 | 15.5 | 22.2 | -89.3271 | 945.25 |
| 188 | 35.16129 | 10.4 | 64.3 | -89.3273 | 965.25 |
| 189 | 35.16129 | 17.7 | 12.6 | -89.3277 | 985.25 |
| 190 | 35.16129 | 6.4 | 54.2 | -89.3281 | 1005.25 |
| 191 | 37.71014 | 12.3 | 11.7 | -88.543 | 865.25 |
| 192 | 37.71014 | 7.6 | 17.5 | -88.5434 | 885.25 |
| 193 | 37.71014 | 16.5 | 13.6 | -88.5435 | 905.25 |
| 194 | 37.71014 | 13.2 | 7.7 | -88.5424 | 925.25 |
| 195 | 37.71014 | 7.6 | 19.1 | -88.5434 | 945.25 |
| 196 | 37.71014 | 8.1 | 59.3 | -88.5433 | 965.25 |
| 197 | 37.71014 | 15 | 24.9 | -88.5434 | 985.25 |
| 198 | 37.71014 | 11.5 | 48.1 | -88.5431 | 1005.25 |
| 199 | 40.03077 | 8.2 | 6.6 | -87.7427 | 865.25 |
| 200 | 40.03077 | 17.4 | 7.2 | -87.7434 | 885.25 |
| 201 | 40.03077 | 12.1 | 16.8 | -87.7433 | 905.25 |
| 202 | 40.03077 | 9.3 | 8.2 | -87.7439 | 925.25 |
| 203 | 40.03077 | 12.1 | 34.1 | -87.743 | 945.25 |
| 204 | 40.03077 | 5.2 | 16.3 | -87.7433 | 965.25 |
| 205 | 40.03077 | 9.9 | 66.6 | -87.743 | 985.25 |
| 206 | 40.03077 | 10.3 | 32.2 | -87.7432 | 1005.25 |
| 207 | 43.95082 | 6.6 | 5.5 | -86.9481 | 865.25 |
| 208 | 43.95082 | 11.7 | 11.6 | -86.9488 | 885.25 |
| 209 | 43.95082 | 14 | 7.3 | -86.9486 | 905.25 |
| 210 | 43.95082 | 5.7 | 31.7 | -86.9491 | 925.25 |

| 211 | 43.95082 | 9.9 | 30.5 | -86.9481 | 945.25 |
| --- | --- | --- | --- | --- | --- |
| 212 | 43.95082 | 11.6 | 36.2 | -86.9482 | 965.25 |
| 213 | 43.95082 | 6.4 | 15.3 | -86.9487 | 985.25 |
| 214 | 43.95082 | 15.2 | 18.4 | -86.9481 | 1005.25 |
| 215 | 47.26667 | 13.5 | 8.2 | -86.1468 | 865.25 |
| 216 | 47.26667 | 5.5 | 9 | -86.1468 | 885.25 |
| 217 | 47.26667 | 10.8 | 7 | -86.1458 | 905.25 |
| 218 | 47.26667 | 12.5 | 21.3 | -86.147 | 925.25 |
| 219 | 47.26667 | 8 | 44.1 | -86.1462 | 945.25 |
| 220 | 47.26667 | 13.5 | 18 | -86.1468 | 965.25 |
| 221 | 47.26667 | 7.2 | 42.1 | -86.1464 | 985.25 |
| 222 | 47.26667 | 6 | 13.5 | -86.1458 | 1005.25 |
| 223 | 9.627907 | 18.6 | 8 | -95.6178 | 875.25 |
| 224 | 9.627907 | 9.8 | 10.4 | -95.6183 | 895.25 |
| 225 | 9.627907 | 17.1 | 20 | -95.6178 | 915.25 |
| 226 | 9.627907 | 5.3 | 35.9 | -95.6178 | 935.25 |
| 227 | 9.627907 | 15.5 | 7.8 | -95.6181 | 955.25 |
| 228 | 9.627907 | 5.1 | 17.2 | -95.6176 | 975.25 |
| 229 | 9.627907 | 10.7 | 58.1 | -95.6175 | 995.25 |
| 230 | 9.065217 | 15.3 | 6.6 | -94.9837 | 875.25 |
| 231 | 9.065217 | 5.2 | 19.3 | -94.9838 | 895.25 |
| 232 | 9.065217 | 11.1 | 6.3 | -94.9839 | 915.25 |
| 233 | 9.065217 | 20.1 | 9.9 | -94.9838 | 935.25 |
| 234 | 9.065217 | 12 | 31.2 | -94.9831 | 955.25 |
| 235 | 9.065217 | 7.4 | 27.8 | -94.9835 | 975.25 |
| 236 | 9.065217 | 7 | 18.8 | -94.9836 | 995.25 |
| 237 | 9.469388 | 15.8 | 15.7 | -94.3494 | 875.25 |
| 238 | 9.469388 | 6.9 | 10.7 | -94.349 | 895.25 |
| 239 | 9.469388 | 18.5 | 17.1 | -94.3493 | 915.25 |
| 240 | 9.469388 | 12.9 | 8.2 | -94.3491 | 935.25 |
| 241 | 9.469388 | 18.7 | 20.8 | -94.3494 | 955.25 |
| 242 | 9.469388 | 6.4 | 56.1 | -94.349 | 975.25 |
| 243 | 9.469388 | 6.2 | 26.9 | -94.3489 | 995.25 |
| 244 | 11.03846 | 17.3 | 10.6 | -93.7128 | 875.25 |
| 245 | 11.03846 | 14 | 5.5 | -93.7131 | 895.25 |
| 246 | 11.03846 | 6 | 15.9 | -93.7127 | 915.25 |
| 247 | 11.03846 | 18.1 | 12.3 | -93.7127 | 935.25 |
| 248 | 11.03846 | 10.2 | 28 | -93.7128 | 955.25 |
| 249 | 11.03846 | 8.1 | 18.1 | -93.713 | 975.25 |
| 250 | 11.03846 | 7.6 | 50.9 | -93.7131 | 995.25 |
| 251 | 11.24138 | 23 | 7 | -93.0786 | 875.25 |
| 252 | 11.24138 | 11.8 | 12.1 | -93.0777 | 895.25 |
| 253 | 11.24138 | 15 | 34 | -93.0781 | 915.25 |
| 254 | 11.24138 | 6.3 | 7.6 | -93.0782 | 935.25 |
| 255 | 11.24138 | 5.1 | 21.6 | -93.0782 | 955.25 |
| 256 | 11.24138 | 10.6 | 20.9 | -93.0784 | 975.25 |
| 257 | 11.24138 | 11.3 | 35.1 | -93.0782 | 995.25 |
| 258 | 10.48276 | 7.4 | 8.4 | -92.4393 | 875.25 |
| 259 | 10.48276 | 18.9 | 8.2 | -92.4392 | 895.25 |
| 260 | 10.48276 | 21.5 | 28.9 | -92.4393 | 915.25 |
| 261 | 10.48276 | 10.8 | 17.4 | -92.4393 | 935.25 |
| 262 | 10.48276 | 4.7 | 18.4 | -92.4391 | 955.25 |
| 263 | 10.48276 | 11.2 | 38.6 | -92.4391 | 975.25 |
| 264 | 10.48276 | 5.3 | 9.3 | -92.4389 | 995.25 |
| 265 | 10.10169 | 11 | 11.5 | -91.7978 | 875.25 |
| 266 | 10.10169 | 6 | 7.6 | -91.7972 | 895.25 |
| 267 | 10.10169 | 5 | 26.5 | -91.797 | 915.25 |
| 268 | 10.10169 | 10.2 | 29.5 | -91.797 | 935.25 |
| 269 | 10.10169 | 7.8 | 10 | -91.7975 | 955.25 |
| 270 | 10.10169 | 8.5 | 22.8 | -91.7972 | 975.25 |
| 271 | 10.10169 | 4.9 | 42 | -91.7972 | 995.25 |
| 272 | 12.82759 | 14.8 | 11.8 | -91.1511 | 875.25 |
| 273 | 12.82759 | 13.6 | 14.9 | -91.1512 | 895.25 |
| 274 | 12.82759 | 7.9 | 14.9 | -91.1508 | 915.25 |
| 275 | 12.82759 | 5.6 | 28.5 | -91.151 | 935.25 |
| 276 | 12.82759 | 10.4 | 46.1 | -91.151 | 955.25 |
| 277 | 12.82759 | 7.6 | 26.7 | -91.1507 | 975.25 |
| 278 | 12.82759 | 17 | 31.4 | -91.151 | 995.25 |
| 279 | 12.50909 | 10.3 | 4.9 | -90.515 | 875.25 |
| 280 | 10.10169 | 8 | 17.4 | -91.7973 | 895.25 |
| 281 | 10.10169 | 15.9 | 36.6 | -91.7974 | 915.25 |
| 282 | 10.10169 | 6.5 | 21.6 | -91.7977 | 935.25 |
| 283 | 10.10169 | 17.7 | 16.7 | -91.7975 | 955.25 |
| 284 | 10.10169 | 14.4 | 21.8 | -91.7973 | 975.25 |
| 285 | 10.10169 | 9.8 | 43.6 | -91.7971 | 995.25 |
| 286 | 14.77193 | 16.3 | 7.3 | -89.869 | 875.25 |
| 287 | 14.77193 | 8.6 | 8.5 | -89.8692 | 895.25 |
| 288 | 14.77193 | 5.3 | 25.5 | -89.8691 | 915.25 |
| 289 | 14.77193 | 8.8 | 23.9 | -89.8695 | 935.25 |
| 290 | 14.77193 | 11.3 | 44 | -89.869 | 955.25 |
| 291 | 14.77193 | 8.3 | 16.4 | -89.8694 | 975.25 |
| 292 | 14.77193 | 13.2 | 53.1 | -89.8689 | 995.25 |
| 293 | 16.70313 | 10.6 | 9.1 | -89.2264 | 875.25 |
| 294 | 16.70313 | 12.9 | 7.9 | -89.2268 | 895.25 |
| 295 | 16.70313 | 17.8 | 16.5 | -89.2269 | 915.25 |
| 296 | 16.70313 | 5.8 | 19.4 | -89.2265 | 935.25 |
| 297 | 16.70313 | 6.7 | 40.4 | -89.227 | 955.25 |
| 298 | 16.70313 | 15 | 37.1 | -89.2265 | 975.25 |
| 299 | 16.70313 | 6.8 | 15.9 | -89.2262 | 995.25 |
| 300 | 17.71014 | 5.9 | 16.6 | -88.5732 | 875.25 |
| 301 | 17.71014 | 16.4 | 12.5 | -88.5734 | 895.25 |
| 302 | 17.71014 | 10.2 | 8.4 | -88.5733 | 915.25 |
| 303 | 17.71014 | 9.9 | 13 | -88.5729 | 935.25 |
| 304 | 17.71014 | 6.5 | 23.3 | -88.573 | 955.25 |
| 305 | 17.71014 | 13 | 59.3 | -88.5727 | 975.25 |
| 306 | 17.71014 | 10.1 | 17.8 | -88.5729 | 995.25 |
| 307 | 20.14925 | 14.4 | 5.2 | -87.9329 | 875.25 |
| 308 | 20.14925 | 8.3 | 21.9 | -87.9333 | 895.25 |
| 309 | 20.14925 | 15.4 | 15.4 | -87.9336 | 915.25 |
| 310 | 20.14925 | 19 | 33 | -87.933 | 935.25 |
| 311 | 20.14925 | 7.4 | 10.7 | -87.9331 | 955.25 |
| 312 | 20.14925 | 7.2 | 31 | -87.9329 | 975.25 |
| 313 | 20.14925 | 15.9 | 24.8 | -87.9332 | 995.25 |
| 314 | 23.04762 | 7.1 | 5.6 | -87.2823 | 875.25 |
| 315 | 23.04762 | 17 | 15.4 | -87.2825 | 895.25 |
| 316 | 23.04762 | 7.6 | 22.9 | -87.2822 | 915.25 |
| 317 | 23.04762 | 17.3 | 25.6 | -87.2828 | 935.25 |
| 318 | 23.04762 | 8.9 | 14.5 | -87.2822 | 955.25 |

| 319 | 23.04762 | 10 | 63.3 | -87.2824 | 975.25 |
| 320 | 23.04762 | 10.4 | 30.3 | -87.2825 | 995.25 |
| 321 | 23.73438 | 13.9 | 8.2 | -86.6288 | 875.25 |
| 322 | 23.73438 | 7.7 | 14.5 | -86.6291 | 895.25 |
| 323 | 23.73438 | 9.9 | 11.6 | -86.6295 | 915.25 |
| 324 | 23.73438 | 12.1 | 39.5 | -86.6291 | 935.25 |
| 325 | 23.73438 | 7.6 | 29 | -86.6289 | 955.25 |
| 326 | 23.73438 | 5.7 | 41.5 | -86.6292 | 975.25 |
| 327 | 23.73438 | 12.4 | 14 | -86.6289 | 995.25 |
| 328 | -10.3721 | 5.4 | 7 | -95.54 | 865.25 |
| 329 | -10.3721 | 19.5 | 8.5 | -95.54 | 885.25 |
| 330 | -10.3721 | 9.2 | 6.2 | -95.54 | 905.25 |
| 331 | -10.3721 | 9.6 | 30.4 | -95.54 | 925.25 |
| 332 | -10.3721 | 12.8 | 23.1 | -95.54 | 945.25 |
| 333 | -10.3721 | 8.8 | 15.4 | -95.54 | 965.25 |
| 334 | -10.3721 | 7.8 | 37.7 | -95.54 | 985.25 |
| 335 | -10.3721 | 5 | 21.3 | -95.54 | 1005.25 |
| 336 | -10.9348 | 10.5 | 11.3 | -94.93 | 865.25 |
| 337 | -10.9348 | 22.3 | 6 | -94.93 | 885.25 |
| 338 | -10.9348 | 23.9 | 15.8 | -94.93 | 905.25 |
| 339 | -10.9348 | 13.7 | 22.1 | -94.93 | 925.25 |
| 340 | -10.9348 | 7.3 | 12.7 | -94.93 | 945.25 |
| 341 | -10.9348 | 5.7 | 43.7 | -94.93 | 965.25 |
| 342 | -10.9348 | 8.3 | 22.9 | -94.93 | 985.25 |
| 343 | -10.9348 | 6.2 | 27.6 | -94.93 | 1005.25 |
| 344 | -10.5306 | 15.7 | 7.1 | -94.32 | 865.25 |
| 345 | -10.5306 | 20.3 | 11 | -94.32 | 885.25 |
| 346 | -10.5306 | 19.1 | 8.4 | -94.32 | 905.25 |
| 347 | -10.5306 | 11.1 | 36.7 | -94.32 | 925.25 |
| 348 | -10.5306 | 9.1 | 11.4 | -94.32 | 945.25 |
| 349 | -10.5306 | 14.4 | 33.6 | -94.32 | 965.25 |
| 350 | -10.5306 | 4.9 | 32.6 | -94.32 | 985.25 |
| 351 | -10.5306 | 9.4 | 25.5 | -94.32 | 1005.25 |
| 352 | -8.96154 | 21.5 | 4.3 | -93.71 | 865.25 |
| 353 | -8.96154 | 6.7 | 10.4 | -93.71 | 885.25 |
| 354 | -8.96154 | 12.9 | 10.3 | -93.71 | 905.25 |
| 355 | -8.96154 | 19.7 | 15.6 | -93.71 | 925.25 |
| 356 | -8.96154 | 9.6 | 13.4 | -93.71 | 945.25 |
| 357 | -8.96154 | 7.2 | 52.5 | -93.71 | 965.25 |
| 358 | -8.96154 | 8.7 | 40.6 | -93.71 | 985.25 |
| 359 | -8.96154 | 6.6 | 17.4 | -93.71 | 1005.25 |
| 360 | -8.75862 | 19.1 | 9.7 | -93.1 | 865.25 |
| 361 | -8.75862 | 25.4 | 7 | -93.1 | 885.25 |
| 362 | -8.75862 | 20 | 17.4 | -93.1 | 905.25 |
| 363 | -8.75862 | 5.5 | 11.2 | -93.1 | 925.25 |
| 364 | -8.75862 | 8.3 | 40 | -93.1 | 945.25 |
| 365 | -8.75862 | 6.3 | 30.2 | -93.1 | 965.25 |
| 366 | -8.75862 | 6.6 | 9.8 | -93.1 | 985.25 |
| 367 | -8.75862 | 11.2 | 31 | -93.1 | 1005.25 |
| 368 | -9.51724 | 7 | 7.3 | -92.49 | 865.25 |
| 369 | -9.51724 | 14.8 | 6.7 | -92.49 | 885.25 |
| 370 | -9.51724 | 17.7 | 22 | -92.49 | 905.25 |
| 371 | -9.51724 | 20.8 | 14.5 | -92.49 | 925.25 |
| 372 | -9.51724 | 11.1 | 31.6 | -92.49 | 945.25 |
| 373 | -9.51724 | 7.6 | 19.7 | -92.49 | 965.25 |
| 374 | -9.51724 | 5.2 | 49.1 | -92.49 | 985.25 |
| 375 | -9.51724 | 5.5 | 9.7 | -92.49 | 1005.25 |
| 376 | -9.77966 | 17.8 | 12.1 | -91.88 | 865.25 |
| 377 | -9.77966 | 12.8 | 8.3 | -91.88 | 885.25 |
| 378 | -9.77966 | 5.3 | 22.8 | -91.88 | 905.25 |
| 379 | -9.77966 | 17.9 | 7.1 | -91.88 | 925.25 |
| 380 | -9.77966 | 11.4 | 21.4 | -91.88 | 945.25 |
| 381 | -9.77966 | 4.6 | 10.6 | -91.88 | 965.25 |
| 382 | -9.77966 | 5.8 | 43.6 | -91.88 | 985.25 |
| 383 | -9.77966 | 8.7 | 34.6 | -91.88 | 1005.25 |
| 384 | -7.50877 | 17.2 | 10.6 | -91.27 | 865.25 |
| 385 | -7.50877 | 8.7 | 13.4 | -91.27 | 885.25 |
| 386 | -7.50877 | 7.2 | 11.1 | -91.27 | 905.25 |
| 387 | -7.50877 | 12.9 | 12.9 | -91.27 | 925.25 |
| 388 | -7.50877 | 10.2 | 46.5 | -91.27 | 945.25 |
| 389 | -7.50877 | 13.9 | 26.1 | -91.27 | 965.25 |
| 390 | -7.50877 | 7.6 | 18.1 | -91.27 | 985.25 |

| 391 | -7.50877 | 8.1  | 41.5 | -91.27 | 1005.25 |
| 392 | -6.81356 | 8.9  | 4.5  | -90.66 | 865.25  |
| 393 | -6.81356 | 10.7 | 21.7 | -90.66 | 885.25  |
| 394 | -6.81356 | 5.9  | 25.9 | -90.66 | 905.25  |
| 395 | -6.81356 | 11.4 | 18.6 | -90.66 | 925.25  |
| 396 | -6.81356 | 5.9  | 24.7 | -90.66 | 945.25  |
| 397 | -6.81356 | 11.3 | 18.8 | -90.66 | 965.25  |
| 398 | -6.81356 | 7    | 10.7 | -90.66 | 985.25  |
| 399 | -6.81356 | 8.3  | 70.5 | -90.66 | 1005.25 |
| 400 | -5.22807 | 13.1 | 7.5  | -90.05 | 865.25  |
| 401 | -5.22807 | 14   | 10.7 | -90.05 | 885.25  |
| 402 | -5.22807 | 18.4 | 23.8 | -90.05 | 905.25  |
| 403 | -5.22807 | 10.8 | 10.7 | -90.05 | 925.25  |
| 404 | -5.22807 | 5.2  | 16.9 | -90.05 | 945.25  |
| 405 | -5.22807 | 7.4  | 28.1 | -90.05 | 965.25  |
| 406 | -5.22807 | 9    | 62.4 | -90.05 | 985.25  |
| 407 | -5.22807 | 13.7 | 26.5 | -90.05 | 1005.25 |
| 408 | -2.91667 | 12.7 | 4.9  | -89.44 | 865.25  |
| 409 | -2.91667 | 8.2  | 18.3 | -89.44 | 885.25  |
| 410 | -2.91667 | 6.4  | 9.5  | -89.44 | 905.25  |
| 411 | -2.91667 | 15   | 16.8 | -89.44 | 925.25  |
| 412 | -2.91667 | 12.4 | 8.7  | -89.44 | 945.25  |
| 413 | -2.91667 | 11.9 | 49.9 | -89.44 | 965.25  |
| 414 | -2.91667 | 13.8 | 17.2 | -89.44 | 985.25  |
| 415 | -2.91667 | 5.7  | 46.3 | -89.44 | 1005.25 |
| 416 | -2.02899 | 5.7  | 4.8  | -88.83 | 865.25  |
| 417 | -2.02899 | 10   | 9.6  | -88.83 | 885.25  |
| 418 | -2.02899 | 10.2 | 31.3 | -88.83 | 905.25  |
| 419 | -2.02899 | 14   | 24.4 | -88.83 | 925.25  |
| 420 | -2.02899 | 5.5  | 25.6 | -88.83 | 945.25  |
| 421 | -2.02899 | 7.9  | 9.8  | -88.83 | 965.25  |
| 422 | -2.02899 | 6    | 39.1 | -88.83 | 985.25  |
| 423 | -2.02899 | 12.7 | 49.9 | -88.83 | 1005.25 |
| 424 | -1.38235 | 14.7 | 8    | -88.22 | 865.25  |
| 425 | -1.38235 | 5.8  | 12.3 | -88.22 | 885.25  |
| 426 | -1.38235 | 12.5 | 13.1 | -88.22 | 905.25  |
| 427 | -1.38235 | 9.1 | 15.1 | -88.22 | 925.25 |
| 428 | -1.38235 | 10.5 | 16.2 | -88.22 | 945.25 |
| 429 | -1.38235 | 15 | 34.9 | -88.22 | 965.25 |
| 430 | -1.38235 | 8 | 53.5 | -88.22 | 985.25 |
| 431 | -1.38235 | 7 | 22.4 | -88.22 | 1005.25 |
| 432 | 0.030769 | 7.6 | 5.7 | -87.61 | 865.25 |
| 433 | 0.030769 | 15.8 | 7.7 | -87.61 | 885.25 |
| 434 | 0.030769 | 9 | 20.5 | -87.61 | 905.25 |
| 435 | 0.030769 | 16 | 16.2 | -87.61 | 925.25 |
| 436 | 0.030769 | 13.2 | 12 | -87.61 | 945.25 |
| 437 | 0.030769 | 16.2 | 55.6 | -87.61 | 965.25 |
| 438 | 0.030769 | 10.5 | 28.1 | -87.61 | 985.25 |
| 439 | 0.030769 | 6.7 | 38.6 | -87.61 | 1005.25 |
| 440 | 3.746032 | 10.2 | 14.2 | -87 | 865.25 |
| 441 | 3.746032 | 8.5 | 5.8 | -87 | 885.25 |
| 442 | 3.746032 | 15 | 12.3 | -87 | 905.25 |
| 443 | 3.746032 | 18.7 | 19.3 | -87 | 925.25 |
| 444 | 3.746032 | 8.6 | 19.9 | -87 | 945.25 |
| 445 | 3.746032 | 6.5 | 21.4 | -87 | 965.25 |
| 446 | 3.746032 | 12.1 | 56 | -87 | 985.25 |
| 447 | 3.746032 | 10.5 | 19.3 | -87 | 1005.25 |
| 448 | -30.3721 | 16.8 | 6.7 | -95.6029 | 875.25 |
| 449 | -30.3721 | 5.5 | 21 | -95.6032 | 895.25 |
| 450 | -30.3721 | 10.5 | 6.7 | -95.6027 | 915.25 |
| 451 | -30.3721 | 21.6 | 11.1 | -95.6035 | 935.25 |
| 452 | -30.3721 | 12.4 | 33.5 | -95.603 | 955.25 |
| 453 | -30.3721 | 7.8 | 29.9 | -95.6029 | 975.25 |
| 454 | -30.3721 | 7.2 | 19.8 | -95.6029 | 995.25 |
| 455 | -30.9348 | 12 | 5.5 | -94.9587 | 875.25 |
| 456 | -30.9348 | 10.4 | 5.8 | -94.9584 | 895.25 |
| 457 | -30.9348 | 11.4 | 23.7 | -94.9587 | 915.25 |
| 458 | -30.9348 | 15.5 | 16.1 | -94.9584 | 935.25 |
| 459 | -30.9348 | 8.6 | 24.2 | -94.9584 | 955.25 |
| 460 | -30.9348 | 6.7 | 8.6 | -94.9586 | 975.25 |
| 461 | -30.9348 | 4.4 | 21.8 | -94.9581 | 995.25 |
| 462 | -30.5306 | 12.4 | 5.8 | -94.3234 | 875.25 |
| 463 | -30.5306 | 12.1 | 6 | -94.3233 | 895.25 |
| 464 | -30.5306 | 11.8 | 24.6 | -94.3235 | 915.25 |
| 465 | -30.5306 | 16.6 | 16.7 | -94.3231 | 935.25 |
| 466 | -30.5306 | 9.1 | 26.1 | -94.3232 | 955.25 |
| 467 | -30.5306 | 7 | 9.4 | -94.3238 | 975.25 |
| 468 | -30.5306 | 4.6 | 23.8 | -94.3233 | 995.25 |
| 469 | -28.9615 | 8 | 8.9 | -93.6883 | 875.25 |
| 470 | -28.9615 | 20.6 | 8.8 | -93.6879 | 895.25 |
| 471 | -28.9615 | 19.3 | 30.3 | -93.6881 | 915.25 |
| 472 | -28.9615 | 11.4 | 18.7 | -93.6883 | 935.25 |
| 473 | -28.9615 | 4.9 | 19.9 | -93.6881 | 955.25 |
| 474 | -28.9615 | 9 | 40 | -93.6876 | 975.25 |
| 475 | -28.9615 | 5.5 | 10.3 | -93.6877 | 995.25 |
| 476 | -28.7586 | 6.2 | 9.8 | -93.0572 | 875.25 |
| 477 | -28.7586 | 24.7 | 13.7 | -93.0568 | 895.25 |
| 478 | -28.7586 | 13.7 | 7.1 | -93.0573 | 915.25 |
| 479 | -28.7586 | 8.5 | 30.6 | -93.0574 | 935.25 |
| 480 | -28.7586 | 13.1 | 15.2 | -93.0567 | 955.25 |
| 481 | -28.7586 | 4.5 | 13.7 | -93.0577 | 975.25 |
| 482 | -28.7586 | 8 | 45.2 | -93.0571 | 995.25 |
| 483 | -29.5172 | 9 | 4.7 | -92.4166 | 875.25 |
| 484 | -29.5172 | 7.4 | 26.2 | -92.4166 | 895.25 |
| 485 | -29.5172 | 23.1 | 17.6 | -92.4166 | 915.25 |
| 486 | -29.5172 | 7 | 27.5 | -92.4167 | 935.25 |
| 487 | -29.5172 | 6 | 11.4 | -92.4168 | 955.25 |
| 488 | -29.5172 | 15.6 | 12 | -92.4165 | 975.25 |
| 489 | -29.5172 | 8.6 | 74.8 | -92.4166 | 995.25 |
| 490 | -29.8983 | 11.7 | 10.1 | -91.7894 | 875.25 |
| 491 | -29.8983 | 19.7 | 11.4 | -91.7894 | 895.25 |
| 492 | -29.8983 | 7.1 | 8 | -91.7895 | 915.25 |
| 493 | -29.8983 | 6 | 18 | -91.7897 | 935.25 |
| 494 | -29.8983 | 11.7 | 36 | -91.7894 | 955.25 |
| 495 | -29.8983 | 16.4 | 12.9 | -91.7894 | 975.25 |
| 496 | -29.8983 | 7.8 | 61.1 | -91.7894 | 995.25 |
| 497 | -27.1724 | 6.5 | 12.5 | -91.1497 | 875.25 |
| 498 | -27.1724 | 18.2 | 9.1 | -91.1499 | 895.25 |
| 499 | -27.1724 | 20.4 | 31.9 | -91.15 | 915.25 |
| 500 | -27.1724 | 8.3 | 26.6 | -91.1502 | 935.25 |
| 501 | -27.1724 | 8.1 | 8.5 | -91.1499 | 955.25 |
| 502 | -27.1724 | 5 | 34.6 | -91.1502 | 975.25 |
| 503 | -27.1724 | 11.6 | 39.1 | -91.1499 | 995.25 |
| 504 | -27.4909 | 17.9 | 9.3 | -90.5184 | 875.25 |
| 505 | -27.4909 | 6.3 | 15.9 | -90.5178 | 895.25 |
| 506 | -27.4909 | 12.1 | 18.8 | -90.5182 | 915.25 |
| 507 | -27.4909 | 7.8 | 8.7 | -90.5185 | 935.25 |
| 508 | -27.4909 | 9.3 | 38.8 | -90.5184 | 955.25 |
| 509 | -27.4909 | 4.8 | 24.7 | -90.5184 | 975.25 |
| 510 | -27.4909 | 13.8 | 36.4 | -90.5181 | 995.25 |
| 511 | -25.2281 | 5.6 | 7.2 | -89.8916 | 875.25 |
| 512 | -25.2281 | 11.1 | 16.9 | -89.8916 | 895.25 |
| 513 | -25.2281 | 10.8 | 11.1 | -89.8916 | 915.25 |
| 514 | -25.2281 | 13.6 | 20.8 | -89.8916 | 935.25 |
| 515 | -25.2281 | 14 | 17.5 | -89.8916 | 955.25 |
| 516 | -25.2281 | 8.7 | 53.3 | -89.8917 | 975.25 |
| 517 | -25.2281 | 5.7 | 25.9 | -89.8917 | 995.25 |
| 518 | -23.2969 | 9.3 | 8.6 | -89.26 | 875.25 |
| 519 | -23.2969 | 6.6 | 24.4 | -89.2601 | 895.25 |
| 520 | -23.2969 | 6.6 | 12.5 | -89.2601 | 915.25 |
| 521 | -23.2969 | 12.5 | 31.7 | -89.26 | 935.25 |
| 522 | -23.2969 | 16.1 | 13.7 | -89.2602 | 955.25 |
| 523 | -23.2969 | 10.3 | 25.7 | -89.2597 | 975.25 |
| 524 | -23.2969 | 6.4 | 40.5 | -89.2599 | 995.25 |
| 525 | -22.3971 | 7.7 | 6 | -88.6229 | 875.25 |
| 526 | -22.3971 | 10.1 | 10 | -88.623 | 895.25 |
| 527 | -22.3971 | 8.4 | 22.1 | -88.6226 | 915.25 |
| 528 | -22.3971 | 7.3 | 37.5 | -88.6226 | 935.25 |
| 529 | -22.3971 | 15 | 25.1 | -88.6229 | 955.25 |
| 530 | -22.3971 | 9.5 | 11.1 | -88.6229 | 975.25 |
| 531 | -22.3971 | 5.8 | 48.9 | -88.6228 | 995.25 |
| 532 | -19.8507 | 11.3 | 10.9 | -87.9992 | 875.25 |
| 533 | -19.8507 | 12.5 | 11.1 | -87.9995 | 895.25 |
| 534 | -19.8507 | 14 | 9.3 | -87.9994 | 915.25 |
| 535 | -19.8507 | 6.8 | 34.4 | -87.9992 | 935.25 |
| 536 | -19.8507 | 20 | 30.1 | -87.9993 | 955.25 |
| 537 | -19.8507 | 5.3 | 19.1 | -87.999 | 975.25 |
| 538 | -19.8507 | 12.8 | 29.2 | -87.9994 | 995.25 |
| 539 | -17.5 | 19.3 | 10.3 | -87.3666 | 875.25 |
| 540 | -17.5 | 7.1 | 14 | -87.3669 | 895.25 |
| 541 | -17.5 | 13.2 | 12 | -87.3672 | 915.25 |
| 542 | -17.5 | 11.1 | 23.1 | -87.3669 | 935.25 |
| 543 | -17.5 | 7.1 | 32.3 | -87.3669 | 955.25 |
| 544 | -17.5 | 18.5 | 43.1 | -87.3673 | 975.25 |
| 545 | -17.5 | 8.2 | 13.1 | -87.3671 | 995.25 |
| 546 | -49.6939 | 22.5 | 5.2 | -94.1419 | 865.25 |
| 547 | -49.6939 | 7.9 | 13 | -94.1421 | 885.25 |
| 548 | -49.6939 | 9.8 | 12.7 | -94.1424 | 905.25 |
| 549 | -49.6939 | 22.4 | 20 | -94.1412 | 925.25 |
| 550 | -49.6939 | 8.9 | 53.7 | -94.1425 | 945.25 |
| 551 | -49.6939 | 10.1 | 8.2 | -94.1423 | 965.25 |
| 552 | -49.6939 | 5.4 | 16.2 | -94.1432 | 985.25 |
| 553 | -49.6939 | 10 | 37.2 | -94.1426 | 1005.25 |
| 554 | -49.4483 | 6 | 7.8 | -93.3481 | 865.25 |
| 555 | -49.4483 | 11 | 12.6 | -93.3489 | 885.25 |
| 556 | -49.4483 | 22.2 | 18.5 | -93.3481 | 905.25 |
| 557 | -49.4483 | 6.4 | 18 | -93.3482 | 925.25 |
| 558 | -49.4483 | 13.6 | 10.7 | -93.348 | 945.25 |
| 559 | -49.4483 | 18 | 12.2 | -93.3483 | 965.25 |
| 560 | -49.4483 | 6.8 | 29.2 | -93.3481 | 985.25 |
| 561 | -49.4483 | 7.5 | 76.3 | -93.3485 | 1005.25 |
| 562 | -48.9483 | 14.3 | 8.4 | -92.5574 | 865.25 |
| 563 | -48.9483 | 13.6 | 6.5 | -92.5576 | 885.25 |
| 564 | -48.9483 | 17 | 24.8 | -92.5572 | 905.25 |
| 565 | -48.9483 | 6.2 | 20.6 | -92.5577 | 925.25 |
| 566 | -48.9483 | 20.8 | 23.9 | -92.5573 | 945.25 |
| 567 | -48.9483 | 10.7 | 14.6 | -92.5574 | 965.25 |
| 568 | -48.9483 | 6.2 | 14.4 | -92.5577 | 985.25 |
| 569 | -48.9483 | 4.5 | 52 | -92.5578 | 1005.25 |
| 570 | -49.8983 | 25.1 | 10 | -91.7639 | 865.25 |
| 571 | -49.8983 | 9.7 | 11.3 | -91.7647 | 885.25 |
| 572 | -49.8983 | 11.1 | 6.6 | -91.7654 | 905.25 |
| 573 | -49.8983 | 5.2 | 28.1 | -91.7642 | 925.25 |
| 574 | -49.8983 | 5 | 14.1 | -91.7649 | 945.25 |
| 575 | -49.8983 | 12.7 | 17.1 | -91.7652 | 965.25 |
| 576 | -49.8983 | 14.4 | 45.4 | -91.7643 | 985.25 |
| 577 | -49.8983 | 5.9 | 28.7 | -91.7641 | 1005.25 |
| 578 | -45.3509 | 7.2 | 6.5 | -90.9759 | 865.25 |
| 579 | -45.3509 | 6.1 | 16.2 | -90.9762 | 885.25 |
| 580 | -45.3509 | 9.6 | 11.9 | -90.9762 | 905.25 |
| 581 | -45.3509 | 16.6 | 22.8 | -90.976 | 925.25 |
| 582 | -45.3509 | 11.7 | 10 | -90.9763 | 945.25 |
| 583 | -45.3509 | 9.6 | 45.6 | -90.9762 | 965.25 |
| 584 | -45.3509 | 7.4 | 23.9 | -90.9758 | 985.25 |
| 585 | -45.3509 | 11.9 | 24.4 | -90.9763 | 1005.25 |
| 586 | -45.1695 | 16.7 | 5.3 | -90.1883 | 865.25 |
| 587 | -45.1695 | 6.4 | 19.2 | -90.188 | 885.25 |
| 588 | -45.1695 | 8.4 | 8.8 | -90.1883 | 905.25 |
| 589 | -45.1695 | 10.5 | 17.4 | -90.1882 | 925.25 |
| 590 | -45.1695 | 16.1 | 28.5 | -90.1877 | 945.25 |
| 591 | -45.1695 | 7 | 24.2 | -90.1884 | 965.25 |
| 592 | -45.1695 | 16.6 | 47.1 | -90.1883 | 985.25 |
| 593 | -45.1695 | 14.4 | 11.6 | -90.1884 | 1005.25 |
| 594 | -42.9167 | 15.2 | 8.6 | -89.4177 | 865.25 |
| 595 | -42.9167 | 12 | 8.7 | -89.4178 | 885.25 |
| 596 | -42.9167 | 7 | 21.2 | -89.4181 | 905.25 |
| 597 | -42.9167 | 7 | 8.7 | -89.4181 | 925.25 |
| 598 | -42.9167 | 6.6 | 20.6 | -89.418 | 945.25 |
| 599 | -42.9167 | 11 | 42 | -89.418 | 965.25 |
| 600 | -42.9167 | 12.5 | 20 | -89.4176 | 985.25 |
| 601 | -42.9167 | 7.1 | 29.9 | -89.4179 | 1005.25 |
| 602 | -42.3971 | 18.4 | 10.3 | -88.6134 | 865.25 |
| 603 | -42.3971 | 7 | 9.3 | -88.6128 | 885.25 |
| 604 | -42.3971 | 7.8 | 15.3 | -88.6131 | 905.25 |
| 605 | -42.3971 | 7.5 | 13.4 | -88.6129 | 925.25 |
| 606 | -42.3971 | 14 | 41.9 | -88.613 | 945.25 |
| 607 | -42.3971 | 12.3 | 13.8 | -88.6135 | 965.25 |
| 608 | -42.3971 | 12.9 | 33.8 | -88.6133 | 985.25 |
| 609 | -42.3971 | 4.8 | 43 | -88.6134 | 1005.25 |
| 610 | -40.9375 | 11.6 | 6.4 | -87.8442 | 865.25 |
| 611 | -40.9375 | 10.4 | 14.2 | -87.8443 | 885.25 |
| 612 | -40.9375 | 15.9 | 14.9 | -87.8442 | 905.25 |
| 613 | -40.9375 | 12.1 | 9.7 | -87.8445 | 925.25 |
| 614 | -40.9375 | 7.1 | 38.3 | -87.8444 | 945.25 |
| 615 | -40.9375 | 6.1 | 20.6 | -87.8444 | 965.25 |
| 616 | -40.9375 | 5.6 | 13.4 | -87.8444 | 985.25 |
| 617 | -40.9375 | 16.2 | 66 | -87.8445 | 1005.25 |
| 618 | -69.5172 | 20 | 12.2 | -92.4188 | 875.25 |
| 619 | -69.5172 | 14.5 | 6.4 | -92.419 | 895.25 |
| 620 | -69.5172 | 6.3 | 21.4 | -92.4192 | 915.25 |
| 621 | -69.5172 | 11.8 | 20.1 | -92.4196 | 935.25 |
| 622 | -69.5172 | 11 | 48.4 | -92.4191 | 955.25 |
| 623 | -69.5172 | 5.9 | 10.3 | -92.4184 | 975.25 |
| 624 | -69.5172 | 14.4 | 20.7 | -92.4193 | 995.25 |
| 625 | -67.5088 | 13.1 | 6.2 | -91.1804 | 875.25 |
| 626 | -67.5088 | 21.7 | 18 | -91.1802 | 895.25 |
| 627 | -67.5088 | 16.5 | 10.2 | -91.1806 | 915.25 |
| 628 | -67.5088 | 5.1 | 24.8 | -91.18 | 935.25 |
| 629 | -67.5088 | 6.9 | 42.1 | -91.1807 | 955.25 |
| 630 | -67.5088 | 13.9 | 32.2 | -91.1807 | 975.25 |
| 631 | -67.5088 | 6.6 | 12.1 | -91.1807 | 995.25 |
| 632 | -65.2281 | 26.2 | 5.1 | -89.9662 | 875.25 |
| 633 | -65.2281 | 8.9 | 23.1 | -89.9663 | 895.25 |
| 634 | -65.2281 | 12.9 | 13 | -89.9661 | 915.25 |
| 635 | -65.2281 | 14 | 7 | -89.9665 | 935.25 |
| 636 | -65.2281 | 14.4 | 34.7 | -89.9657 | 955.25 |
| 637 | -65.2281 | 6.6 | 15.4 | -89.9659 | 975.25 |
| 638 | -65.2281 | 9.1 | 64.7 | -89.9671 | 995.25 |
| 639 | -62.029 | 20.9 | 14.4 | -88.7649 | 875.25 |
| 640 | -62.029 | 22.9 | 7 | -88.7646 | 895.25 |
| 641 | -62.029 | 7.3 | 13.9 | -88.7656 | 915.25 |
| 642 | -62.029 | 7.5 | 14.1 | -88.766 | 935.25 |

| 643 | -62.029 | 6.3 | 37.4 | -88.765 | 955.25 |
| 644 | -62.029 | 17.3 | 46.7 | -88.7657 | 975.25 |
| 645 | -62.029 | 9.6 | 16.8 | -88.7653 | 995.25 |